# Simulation of environmental impacts on the synthesis of carbyne with more than 6000 atoms for emerging continuously tunable energy barriers in CNT-based transistors

[1,2]Chi Ho Wong*,[5]Yan Ming Yeung, [4]Xin Zhao, [1]Wing Cheung Law, [1]Chak-yin Tang, [6]Chee Leung Mak, [6]Chi Wah Leung,[7]Lei Shi, [3]Rolf Lortz

[1]*Department of Industrial and Systems Engineering, The Hong Kong Polytechnic University, Hong Kong, China*

[2]*Research Institute for Advanced Manufacturing, The Hong Kong Polytechnic University, Hong Kong, China*

[3]*Department of Physics, The Hong Kong University of Science and Technology, Hong Kong, China*

[4]*Department of Biomedical Engineering, The Hong Kong Polytechnic University, Hong Kong, China*

[5]*School of Science, The Hong Kong University of Science and Technology, Hong Kong, China*

[6]*Department of Applied Physics, The Hong Kong Polytechnic University, Hong Kong, China*

[7]*State Key Laboratory of Optoelectronic Materials and Technologies, Nanotechnology Research Center, School of Materials Science and Engineering, Sun Yat-sen University, Guangzhou, 510275, China*

*Email: roych.wong@polyu.edu.hk*

**Abstract:**

Transistors made up of carbon nanotubes CNT have demonstrated excellent current-voltage characteristics which outperform some high-grade silicon-based transistors. A continuously tunable energy barrier across semiconductor interfaces is desired to make the CNT-based transistors more robust. Despite the direct band gap of carbyne inside a CNT can be widely tuned by strain, the size of carbyne cannot be controlled easily. The production of a monoatomic chain with more than 6000 carbon atoms is an enormous technological challenge. To predict the optimal chain length of a carbyne in different molecular environments, we have developed a Monte Carlo model in which a finite-length carbyne with a size of 4000-15000 atoms is encapsulated by a CNT at finite temperatures. Our simulation shows that the stability of the carbyne@nanotube is strongly influenced by the nature and porosity of the CNT, the external pressure, the temperature and the chain length. We have observed an initiation of chain-breaking process in a compressed carbyne@nanotube. Our work provides much needed input for optimising the carbyne length to produce carbon chains much longer than 6000 atoms at ~300K. Design rules are proposed for synthesizing ~1% strained carbyne@(6,5)CNT as a component in CNT-based transistors to tune the energy barriers continuously.

## 1. Introduction:

Continuously adjusting the band gap of semiconductors is expected to revolutionize semiconductor technology [1, 2]. The charge current flowing through semiconductor devices such as p-n junctions, field-effect transistors, bipolar junction transistors are related to the energy barrier across the transition region between semiconductor interfaces [2]. Recently, carbon nanotube (CNT) transistors outperform some high-class silicon-based transistors and [3] this remarkable performance deserves attentions. If the band gap of a CNT can be tuned continuously, more robust CNT transistors will debut and it will diversify the applications of CNT transistors. However, tuning the band gap of semiconductors continuously [1], especially for carbon nanotube, is an uphill challenge. The band gap of carbon nanotube depends on discrete chirality numbers [1]. In order to get out of this dilemma, a continuously tunable length-dependent direct band gap of carbyne (in form of a linear carbon chain LCC) has been proposed for fine-tuning the energy barriers across semiconductor junctions [1,4]. Carbyne-based transistors have been demonstrated an outstanding current-voltage IV characteristic experimentally and the IV performance always depends on the applied strain or the size of carbyne [5], in which the band gap of carbyne can also be widely tuned by applying less than 10% strain [6]. The strain or length-triggered continuously tunable band gap of the carbyne inside a CNT may give a solution to widening the practical uses of CNT-based transistors. Unfortunately, the length of carbyne during sample fabrication is unpredictable even though advanced manufacturing techniques applied. There is no systematic routine that provides the theoretical input for optimising the chain length of carbyne, which hinders experimental bottom-up studies of the continuously tunable band gap.

The preparation of carbon chains with a length in the order of 20 atoms has been a difficult problem in the last decade [7,8]. To extend the chain length, Shi *et al.* encapsulated the carbyne with carbon nanotubes (CNTs) as a nanoreactor, resulting in carbon chains with up to 6000 atoms [9]. Based on the analysis of the carbyne@nanotube, it was argued that the Van der Waal's (VDW) force emanating from the carbon nanotube plays an important role in stabilising the internal carbyne [4,9]. Inspired by the work of Shi *et al.*, Chi Ho Wong *et al.* developed a Monte Carlo method (Linear Carbon Chains (LCC) model) to study the stability of free-standing carbon chains laterally coupled by VDW forces at finite temperatures [10]. In the absence of carbon nanotubes, the direct action of the environment on the carbyne can be detrimental to the development of long chain lengths, but the kink structure [11, 12] in short carbyne can induce strong ferromagnetism [13]. One year later, Chi Ho Wong *et al.* succeeded in fabricating a ferromagnetic VDW-coupled free-standing carbon chain array with a length of ~100 atoms at room temperature assisted by the LCC model [13], where its ferromagnetism survives above 400K. This observation is consistent with the argument of Shi *et al.* that the VDW interaction can be used to stabilise carbyne. On the other hand, several dopants have been proposed to enhance the magnetism of carbyne-based semiconductors with a local magnetic moment above $1.7\mu_B$ [13,14].

The discovery of the 400K ferromagnetism makes carbyne becomes a potential material for room-temperature spintronic applications down to monoatomic scale. To take spintronic nanodevices to the next level, edge magnetism [15] of semiconductors is the most promising route. Fortunately, the truly one-dimensional monoatomic magnetic semiconductor (i.e. carbyne) with exceptionally mechanical strength already consists of an edge everywhere. The monoatomic structure makes carbyne an ideal candidate for studying the physics of edge magnetism, as it is much thinner than any unit cell thick nanowire or ultrathin zigzag nanoribbons of graphene [16,17]. From scientific point of view, the spin-spin interaction of a monoatomic chain contributes to one of the types of boundary magnetism in a higher-dimensional Bravais lattice. If the chain length of magnetic carbyne can be properly tuned [11-13], the length-dependent spin-spin interaction can be experimentally investigated down to the monoatomic scale, providing not only valuable scientific information for monitoring the interfaces of magnetic compounds for spintronics [18], but also insights into the interface problems of magnetic heterostructures [19] on an industrial scale.

In spite of the technological and scientific advances, a tiny change in experimental parameters can fragmentize carbyne during sample fabrications. Theoretical guidelines are needed to control the carbyne length within a targeted range. However, the LCC model can only compare the stability of fixed-length carbyne [10], which is not designed for a variable-length condition. The kink structuring carbyne chain simulated by the LCC model remains continuous no matter how long the chain is. A similar problem can be noticed in DFT software that chain breakage is not observed in an isolated carbyne even though the carbyne is very long, and the entire isolated chain remains linear surprisingly after applying geometric optimizations. Worse still, the DFT study of finite length carbyne is limited to only ~100-200 atoms due to prohibitive computational cost [20]. For example, a ~6000 atoms long carbyne surrounded by CNT (over ~100000 atoms per unit cell) may not be a realistic task for DFT software [21]. To speed up the computational progress, using DFT software to calculate an infinitely long LCC with few atoms per unit cell is inevitably a popular option [14,22]. DFT methods are ideal for situations at 0 K [23]. DFT results are usually representative unless the materials are extremely unstable at finite temperatures. If a long linear carbon chain (LCC) dissociates at room temperature, a transfer function should be imposed on DFT methods to align theoretical results with experimental observations. A better software is needed to monitor carbyne in variable-length condition. Otherwise, it is unclear how sample quality affects the growth of internal LCC, and how the fragmentation of LCC takes place.

In view of the above problems, we developed a Monte Carlo algorithm to predict the stability of carbyne@nanotubes at finite temperatures with reasonable computational effort. A stochastic process is monitored to predict the stability of the carbyne@nanotubes up to 300 K. Parametric studies of the composite focus on the external pressure, temperature, chain length of carbyne, chirality and porosity of the carbon nanotube, etc. Molecular modelling for the formation of fragmentized carbyne will be debut in this work. After the Monte Carlo algorithm is introduced, we will propose a potential LCC@CNT composite to vary the band gap continuously in the CNT-based transistors.

## 2. Methods:

### 2.1. Hamiltonian

The Hamiltonian of the carbyne@nanotube with length L is

$$H = e^{-T/T_{bj}} \left( \sum_{n=1,3,5}^{N} |E_{n,j} - E_1| e^{-\frac{\ell_n - \ell_{n,j}^{eq}}{0.5\ell_{n,j}^{eq}}} + \sum_{n=2,4,6}^{N} |E_{n,j'} - E_3| e^{-\frac{\ell_n - \ell_{n,j'}^{eq}}{0.5\ell_{n,j'}^{eq}}} \right) + e^{-T/T_{bj}} \sum_{n=1,2,3}^{N} J_A e^{-\frac{\ell_n - \ell_{n,j}^{eq}}{0.5\ell_{n,j}^{eq}}} \left(\cos\theta + 1\right)^2$$

$$-4\varphi \sum_{\phi=0}^{2\pi} \sum_{n=1}^{N} \left[ \left(\frac{\sigma}{r}\right)^6 - \left(\frac{\sigma}{r}\right)^{12} \right]$$

where N is the total number of atoms in the carbon chain and T is the surrounding temperature. The $C-C$, $C=C$ and $C \equiv C$ bond energies are $E_1$ = 348kJ/mol, $E_2$ = 614 kJ/mol and $E_3$ = 839 kJ/mol, respectively [10]. The type of covalent bond is proposed by stochastic variables j and j'. For example, $E_{n,j}$(n = 100, j = 3) refers to the 100$^{th}$ carbon atom (n = 100) forming a triple bond (j = 3) with respect to the 99$^{th}$ carbon atom (n-1 = 99). The temperature of bond dissociation is $T_{bj} = E_j/k_B$, where $k_B$ is the Boltzmann constant [26]. The chain-stability factor is defined by $e^{-\frac{\ell_n - \ell_{n,j}^{eq}}{0.5\ell_{n,j}^{eq}}}$ [10].

The bond distance $\ell$ is computed in Cartesian coordinates. The longitudinal axis refers to x-axis while y- and z-axis form a lateral plane. The $C-C$, $C=C$ and $C \equiv C$ bond length on the ground state are $\ell_{n,1}^{eq}$ = 154pm, $\ell_{n,2}^{eq}$ = 134pm and $\ell_{n,3}^{eq}$ = 120pm, respectively [10]. For the kink term, the bond angle (or pivot angle) between three adjacent carbon atoms is θ. A linear carbon chain means θ = 180 degrees, where the angular energy $J_A$ is ~600 kJ/mol [10]. The Van der Waal's interaction between the carbon nanotube and the internal chain is $E_{vdw} = -4\varphi \sum_{\phi=0}^{2\pi} \sum_{n=1}^{N} \left[ \left(\frac{\sigma}{r}\right)^6 - \left(\frac{\sigma}{r}\right)^{12} \right]$ [10]. The $\sum_{\phi=0}^{2\pi}$ sums up VDW interaction along the angular plane ϕ of a CNT and r is the radial separation between the LCC and the CNT. By considering $L\frac{dE_{vdw}^2}{dL^2} = \frac{1}{\zeta}$ and $\frac{\partial E_{vdw}}{\partial r} = 0$ [10], the calculated VDW constants are $\sigma$ ~

$1.2 \times 10^{-10}$m and $\varphi \sim 8 \times 10^{-23}$J [10] where $\zeta$ is the isothermal compressibility. VDW interaction is the only coupling between the LCC and the CNT.

## 2.2. Initial condition

A 6000-atom-long LCC (unless otherwise specified) surrounded by a single-walled carbon nanotube (SWCNT) is studied, which is abbreviated as LCC@($N_c$,$M_c$)CNT as shown in Figure 1a. The LCC@($N_c$,$M_c$)CNTs are spaced by ~0.5nm laterally in an parallel array, where ($N_c$,$M_c$) refers to the chirality of CNT. After applying geometric optimization to a periodic LCC@($N_c$,$M_c$)CNT at the GGA level under the LAPW method (Muffin radius ~1.3au; the total number of k-points in the Brillouin zone ~ 200) [24,25], the repeated unit of CNT is imported in our Monte Carlo simulation. The initial bond length of the internal LCC (cumulene phase: consecutive double bond) is 134pm. The lengths of the LCC and ($N_c$,$M_c$)CNT are equal. The length of the ($N_c$,$M_c$)CNT in our Monte Carlo simulation can be scaled using the repeated unit. The porous carbyne@nanotube and the radically compressed carbyne@nanotube are drawn schematically in Figure 1b and Figure 1c, respectively.

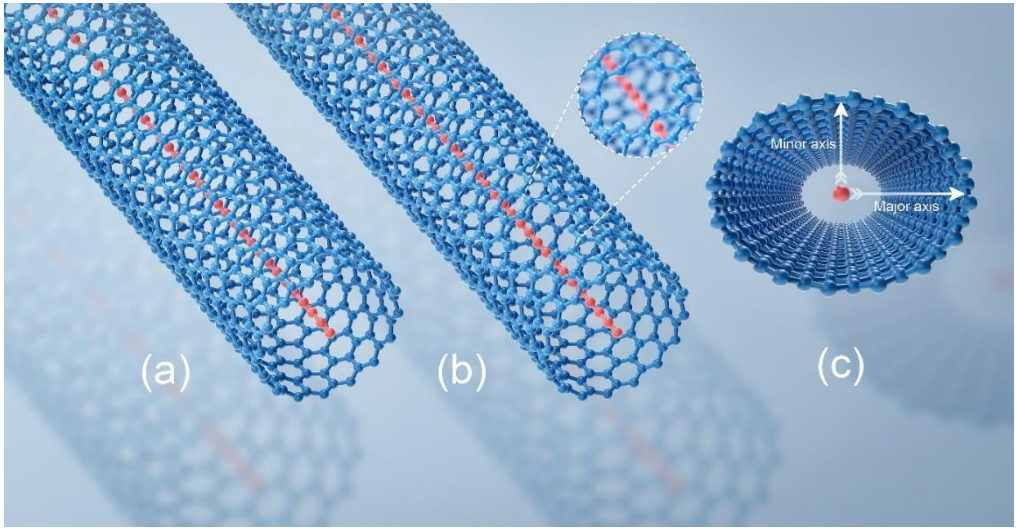

Figure 1: The carbon nanotube is marked by blue color. The red internal chain refers to carbyne. **a** carbyne@nanotube. **b** porous carbyne@nanotube. **c** the cross section of a radially compressed carbyne@nanotube. The compression is along the minor axis.

## 2.3. Algorithm

During the Monte Carlo iterations, the atomic coordinate and the type of covalent bond in the carbon chain are amended at finite temperatures. In each Monte Carlo step (MCS), we randomly select an atom in the carbon chain and then calculate the initial Hamiltonian.

To propose the trial types of covalent bonds at each MCS, we assign a random number $0 \leq R_{bond} \leq 1$ in Table 1. The proposed types of covalent bond depend on the value of $R_{bond}$. For example, the selected $n^{th}$ atom forming two double bonds with its nearest neighbours is expressed as $[=C=]$; If $[=C=]$ is detected and $R_{bond} = 0.8$, the trial type of covalent bond is $[-C-]$.

Table 1: The trial state of a covalent bond.

| (A) If $[=C=]$ is detected | Trial state |
|---|---|
| $0 \leq R_{bond} < 0.33$ | $[\equiv C-]$ or $[-C\equiv]$ in equal probability |
| $0.33 \leq R_{bond} < 0.66$ | $[=C-]$ or $[-C=]$ in equal probability |
| $0.66 \leq R_{bond} \leq 1$ | $[-C-]$ |
| (B) If $[\equiv C-]$ or $[-C\equiv]$ is detected | Trial state |
| $0 \leq R_{bond} < 0.33$ | $[=C=]$ |
| $0.33 \leq R_{bond} < 0.66$ | $[=C-]$ or $[-C=]$ in equal probability |
| $0.66 \leq R_{bond} \leq 1$ | $[-C-]$ |
| (C) If $[-C-]$ is detected | Trial state |
| $0 \leq R_{bond} < 0.33$ | $[\equiv C-]$ or $[-C\equiv]$ in equal probability |
| $0.33 \leq R_{bond} < 0.66$ | $[=C-]$ or $[-C=]$ in equal probability |
| $0.66 \leq R_{bond} \leq 1$ | $[=C=]$ |
| (D) If $[=C-]$ or $[-C=]$ is detected | Trial state |
| $0 \leq R_{bond} < 0.33$ | $[\equiv C-]$ or $[-C\equiv]$ in equal probability |
| $0.33 \leq R_{bond} < 0.66$ | $[-C-]$ |
| $0.66 \leq R_{bond} \leq 1$ | $[=C=]$ |

Based on the metropolis algorithm [26] and the LCC model [10], we propose a trial range for the spatial fluctuation at each MCS. Three random numbers $R_x$, $R_y$, $R_z$ from 0 to 1 are generated, respectively. If $R_x > 0.5$, the selected atom moves longitudinally in the +x direction. Otherwise, it moves in the -x direction. The same cut-off value of 0.5 is also applied to the sign convention along y- and z- axis. After the trial directions are proposed, the trial fluctuations (δx, δy, δz) of the carbon atom are $\delta x = \pm v < \delta t > R_c$ and $\delta y = \delta z \sim \frac{k_B T}{< E_1 + E_2 + E_3 >} \delta x$, where the free particle velocity is $v = \sqrt{\frac{k_B T}{M}}$ [27]. The average scattering time $<\delta t>$ and the mass of a carbon atom M are ~$10^{-13}$s and $19.9 \times 10^{-27}$kg, respectively [10]. The random number $0 < R_c < 1$ is to fine-tune the spatial fluctuation at each MCS.

The Hooke's factor, $f(Hooke) = \frac{K_{break}(\ell_{max} - <\ell_{eq}>)}{K_{carbyne@CNT}(\ell_{carbyne@CNT}|_{0K} - <\ell_{eq}>)}$, acts as a mean-free-path MFP controller to adjust the relative dynamics ($\delta x \cdot f(Hooke)$, $\delta y \cdot f(Hooke)$, $\delta z \cdot f(Hooke)$) from a chain-breaking state to a crystalline form. The Hooke's factor mimics the situation where a stable carbyne@nanotube should have a shorter MFP and narrower range of spatial fluctuation than an unstable one. The numerator of $f(Hooke)$ marks the elastic force when an isolated LCC is elongated from a ground state to a break point [27]. The break point refers to the formation of the maximum $C-C$ length $\ell_{max} \sim 1.73$ Å where the MFP is greatest [28]. The ground state of an isolated LCC (cumulene phase) has an average bond length of $<\ell_{eq}> \sim 1.34$Å. The spring constant of a LCC at the break point is $K_{break}$. The denominator of $f(Hooke)$ monitors the elastic force of a very long LCC protected by a CNT relative to the same $<\ell_{eq}>$ [27]. The spring constant and average bond length of the internal LCC at 0K is $K_{carbyne@CNT}$ and $\ell_{carbyne@CNT}|_{0K}$, respectively. In our DFT simulation, the

GGA-PBE functional is chosen [24,25]. If we set the bond length to be 0.173nm, we obtain the $K_{break}$ of an isolated infinitely long LCC chain. The $K_{carbyne@CNT}$ refers only to the spring constant of the internal LCC along the chain axis.

After the trial range of spatial fluctuations and trial types of covalent bonds are proposed, we estimate the trial Hamiltonian. If the trial Hamiltonian is less positive (or more negative) than the initial Hamiltonian, the trial states are accepted [26]. Otherwise, the system returns to the previous states [26]. Thermal excitation is another opportunity to accept or reject the trial states in parallel. If the random number $0 \leq R_B \leq 1$ is smaller than the Boltzmann factor $e^{\frac{-\Delta H}{k_B T}}$, the trial states are accepted [26]. Otherwise, the system discards the trial states. The Monte Carlo process continues until the Hamiltonian vs Monte Carlo steps approach a flat line (~250000 steps). We average the data in the range of $230000 \leq MCS \leq 250000$. We impose a boundary condition that the location of the 1st atom in LCC is fixed and the covalent bond between the 1st and 2nd atoms in LCC is always $C-C$ bond. The initial condition is executed again at each temperature. The octet rule [27] is applied.

### 2.4: Pre-calibration

The average bond distance of a finite length LCC in the CNT $\ell_{carbyne@CNT}|_{0K}$ is unknown before the simulation begins. To determine a reasonable trial value of $\ell_{carbyne@CNT}|_{0K}$, we use the recently presented LCC model [10] to generate a long LCC of known N. We calculate the average bond distance of the LCC $\ell_{carbyne}|_{501K}$ above the Peierls transition temperature at ~500K [10,22], at which the polyyne phase forms automatically. Since $K_{carbyne@CNT}$ and $<\ell_{eq}>$ refer to the situation at 0K, the $\ell_{carbyne}|_{0K}$ is estimated with the help of $\frac{\ell_{carbyne}|_{501K} - \ell_{carbyne}|_{0K}}{\ell_{carbyne}|_{0K}} = \alpha \Delta T$, where the thermal expansion coefficient is $\alpha = 7 \times 10^{-5} K^{-1}$ [10,29]

To obtain the value of $\ell_{carbyne@CNT}|_{0K}$, we perform a pre-calibration by substituting the trial $\ell_{carbyne}|_{0K}$ into the Hooke's factor. Then we perform the Monte Carlo simulation (Section 2.2-2.3) of LCC@($N_c$,$M_c$)CNT at T ~ 0K with the trial Hooke's factor. During the Monte Carlo iterations of the internal LCC, the effect of CNT tunes the average bond length from $\ell_{carbyne}|_{0K}$ to $\ell_{carbyne@CNT}|_{0K}$. After the pre-calibration, we substitute $\ell_{carbyne@CNT}|_{0K}$ to the Hooke's factor. With the calibrated Hooke's factor, our simulation tool is ready to study the stability of LCC@($N_c$,$M_c$)CNT.

### 3. Results:

Thermal excitation from 0K to 300K has changed the proportion of polyyne and cumulene phase in Figure 2. We found that the probability of forming a $C_{1-3}$ bond (polyyne phase) in the internal LCC decreases from 1.00 to 0.96 upon heating. Due to the growth of the polyyne phase, the probability of forming a $C_{2-2}$ bond (cumulene) correspondingly increases to ~0.04. The formation of $C_{1-2}$ ($[=C-]$ or $[-C=]$) and $C_{1-1}$ ($[-C-]$) bonds is rare. The most energetically favourable phase of the internal LCC is still polyyne, even when temperature is 300K [9]. After the pre-calibration process at T ~ 0K, the average bond length of the internal LCC is ~1.3Å and the bond-length alternation is above 99.9%.

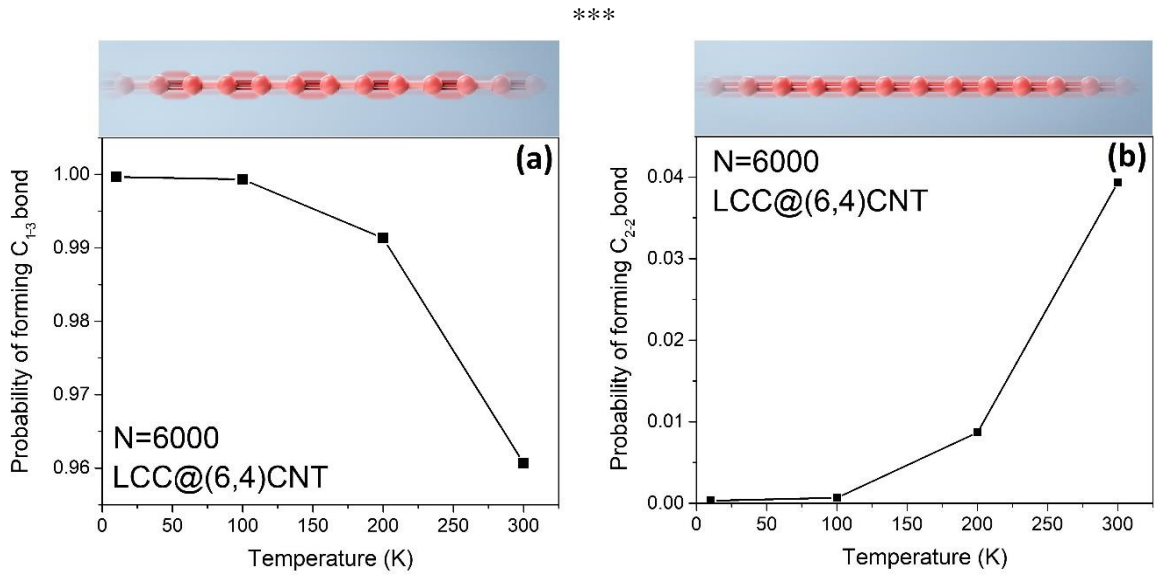

Figure 2: A LCC encapsulated by a (6,4)CNT upon heating. The LCC is made up of 6000 atoms. **a** The probability of maintaining a polyyne phase in the carbon chain. The $C_{1-3}$ bond is defined as the formation of single bond and triple bond alternatingly. The red chain shows the structure of polyyne schematically. **b** The probability of maintaining a cumulene phase in the carbon chain. The $C_{2-2}$ bond is defined as the formation of consecutive double bonds. The red chain refers to the schematic structure of cumulene.

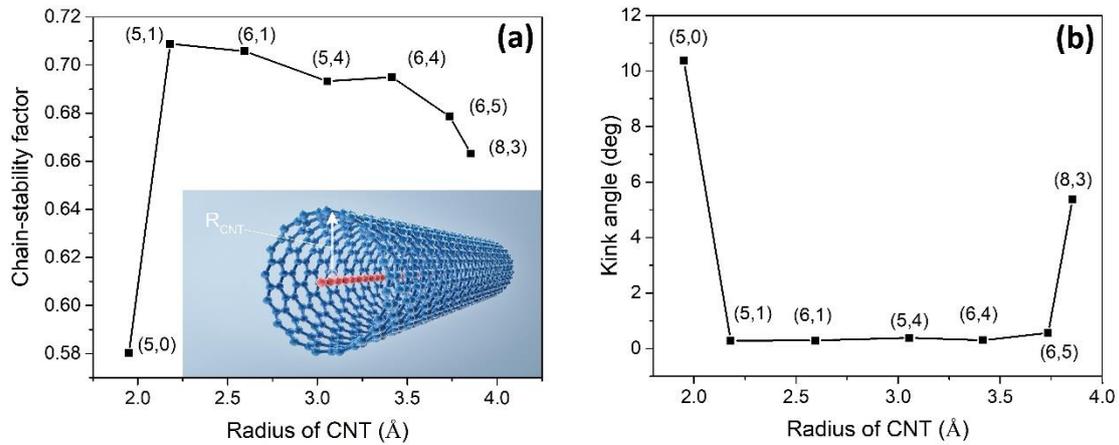

Figure 3: **a** The stability of the 6000-atom-long carbyne@nanotube depends on the radius of carbon nanotube $R_{CNT}$ and the chirality $(N_c, M_c)$. The red chain is LCC while the blue tube is CNT. The temperature of the thermal bath is 300K. For the chain stability factors near 0.7, the average bond lengths of the LCCs range from ~1.3 to ~1.4Å. **b** The average kink angles of the internal LCCs are plotted accordingly. A kink angle of zero (or pivot angle =180 deg) means that the chain is linear. The standard deviation of the LCC@(6,4)CNT is as low as ~0.1.

As illustrated in Figure 3a, the stability of the carbyne@nanotube depends on the radius of the CNT associated with its chirality $(N_c, M_c)$ [4,9]. In the carbyne@nanotube experiment [9], (6,4) and (6,5) CNTs were the most abundant components. At a constant temperature of 300K, our simulation data shows that the (6,4)CNT should have a longer LCC than (6,5)CNT. Although the experiment [9] does not contain (5,0), (5,1), (6,1) and (5,4)CNTs, we used these samples in our simulation. The

LCC@(5,1)CNT have the highest chain stability factor among the others, whereas the chain-stability of LCC@(5,0)CNT is the worst. Figure 3b demonstrates how the kink angles of the internal LCC are affected by the geometry of the CNTs. The average kink angle of LCC@(5,0)CNT is much higher than the others. It can be observed that the higher the chain stability factor, the lower the average kink angle of the LCC.

The chain stability of the internal LCC consisting of 6000 atoms and 15000 atoms is reduced by thermal excitation, as demonstrated in Figure 4. The LCC with 6000 atoms in length has a higher chain stability factor $e^{-\frac{\ell_n - \ell_{n,j}^{eq}}{0.5 \ell_{n,j}^{eq}}}$ than the LCC with 15000 atoms in length, regardless of the temperature. Comparing the chain stabilities in both cases, it can be interpreted that the atomic fluctuations of the 6000-atom-long LCC is always lower at the same temperature. The chain-stability of the carbyne@nanotube in different lengths is compared in Figure 5. At T=300K, the chain stability factor gradually decreases when N is less than 5500. However, at N > 6000, a rapid reduction in the chain stability factor is observed when we halve the cut-off length of the LCC to be N = 5750. The inset of Figure 5 refers to the average energy of the internal LCC as a function of the Monte Carlo iterations.

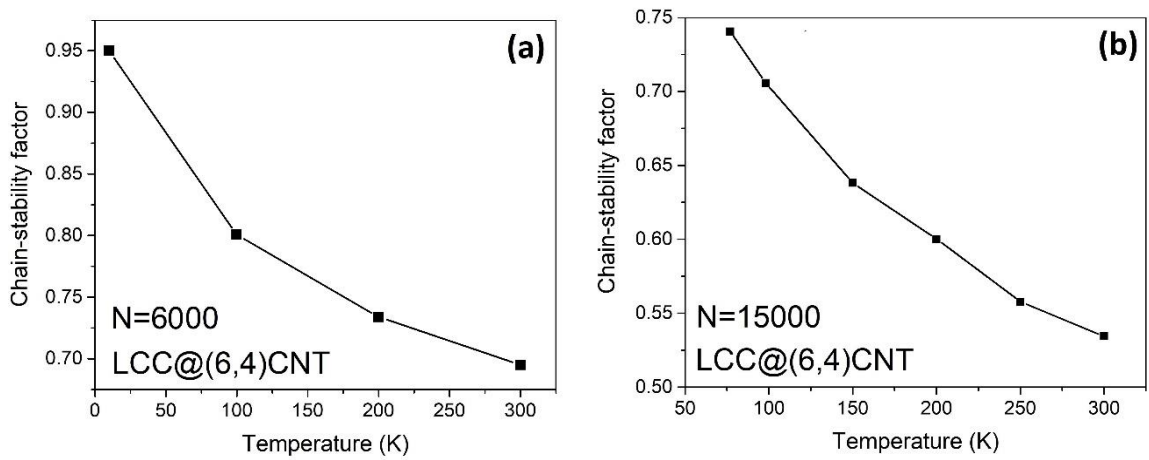

Figure 4: A LCC encapsulated by a (6,4)CNT during heating. The atomic fluctuations of the LCC depend on the chain length and temperature. **a** The LCC consists of 6000 atoms. **b** The LCC contains 15000 atoms.

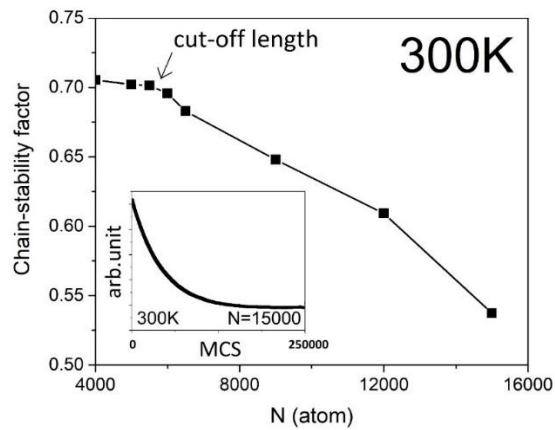

Figure 5: The chain-stability factor is a measure of atomic fluctuations of LCC@(6,4)CNT. The reduction in the chain stability factor is more pronounced at N > 6000. The inset shows that our sample reached an equilibrium state at the Monte Carlo steps exceeding ~180000.

A compressive strain in the radial direction to the LCC@(6,4)CNT deteriorates the chain stability. The chain stability factor drops from 0.69 to 0.68 at a compressive strain of 2%. However, when the compressive strain reaches 4% in test 1 (Table 2), the formation of seven extremely sharp pivot angles within ~10° < pivot angle < ~60° at n = 103$^{rd}$, 576$^{th}$, 932$^{nd}$, 1853$^{rd}$, 4318$^{th}$, 4943$^{rd}$, 5705$^{th}$ atom are observed in Figure 6. The critical kink structure around n = 4943$^{rd}$ (marked by red circle) is drawn. The carbon-carbon distances nearby the sharp kink points always reach ~1.73Å. The carbon atoms at the sharp kink points are mostly connected by $C-C-C$. More samplings are shown in Table 2 where we repeat the studies of 4% radial compression in test 2 and test 3.

The porosity of the (6,4)CNT harms the chain stability of the internal LCC as demonstrated in Table 3. The introduction of a tiny amount of vacancy defects is sufficient to significantly shorten the cut-off length to N ~ 5600. In Table 4, the cut-off length of the LCC@($N_c$,$M_c$)CNT in various chiralities at 300K is investigated. The internal LCC in the (5,1)CNT is the longest. The internal LCC in the (6,4)CNT is ~10% longer than that in a (6,5)CNT. Based on the data in Table 5, it is possible to extend the internal LCC to a length of 15000 atoms. However, the sample should be cooled to T~120K. Our model expects that the internal LCC in the (6,4)CNT should increase by less than 10% at 273K.

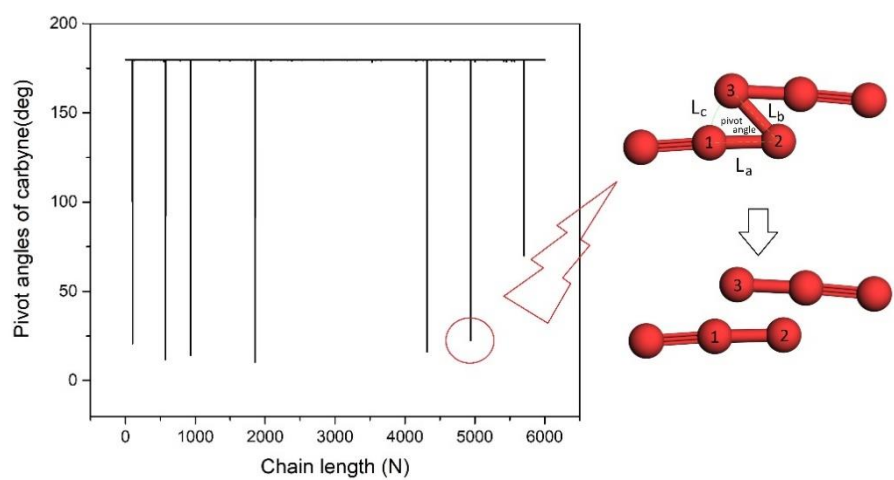

Figure 6: The pivot-angle distribution of the carbon chain enclosed by a ~4% radially compressed CNT. Seven critical kink structures with shape pivot angles are observed. One of the critical kink structures are drawn schematically (not-to-scale). $L_c$, $L_b$, and $L_a$ forms a pivot angle. $L_b$ ~ 1.73Å and 1.5Å < $L_a$ < 1.6Å are always observed in all the critical kink structures. Formation of 7+1 = 8 carbyne fragments are expected at an instantaneous moment of chain breakage in Test 1 (Table 2). The electrostatic repulsion between the sites '1' and '3' is stronger than triple-bond strength.

Table 2: The probable sizes of carbyne fragments in the ~4% compressed (6,4)CNT at 300K

| Samplings | The sizes of carbyne fragments (atoms) |
|---|---|
| Test 1 | 102, 473, 365, 921, 2465, 625, 762, 287 |
| Test 2 | 546, 1650, 2251, 947, 138, 468 |
| Test 3 | 71, 755, 164, 2003, 1081, 949, 207, 501, 269 |

Table 3: Effect of vacancy defects on the stability of the LCC@(6,4)CNT at 300K

| Vacancy Defect (%) | Cut-off length of LCC@ (6,4)CNT |
|---|---|
| 0 | 5750-atom-long |
| 2.5 | 5600-atom-long |

Table 4: Prediction of the LCC lengths in different carbyne@nanotubes at 300K.

| Chirality (Nc,Mc) | Cut-off length of LCC@ (Nc,Mc)CNT |
|---|---|
| (8,3) | 4850-atom-long |
| (6,5) | 5250-atom-long |
| (6,4) | 5750-atom-long |
| (5,1) | 7200-atom-long |

Table 5: The predicted cut-off lengths of the LCC in a (6,4)CNT at different temperatures.

| Cut-off length of LCC | Operating temperature (K) |
|---|---|
| 5750-atom-long | 300 |
| 6600-atom-long | 273 |
| 15000-atom-long | 122 |

## 4. Discussion:

Although the cumulene phase should dominate in an isolated LCC [22], the polyyne phase is energetically more favourable in the LCC@($N_c$,$M_c$)CNT [9]. According to Figure 2, the polyyne phase fades upon heating because the atomic fluctuations are more severe at high temperatures. This process can be controlled by the Boltzmann factor [26] which more readily accepts the trial type of covalent bonds at high temperatures [10]. The formation of $[-C=]$, $[=C-]$ and $[-C-]$ is rare, confirming that our Monte Carlo simulation effectively monitors the process of energy minimization. For instance, the energy difference between $[-C\equiv]$ and $[=C=]$ is $|348 + 839 - (614 + 614)| = 41$ kJ/mol per site. However, the energy difference (per site) between $[-C\equiv]$ and $[-C=]$, and the energy difference (per site) between $[-C\equiv]$ and $[-C-]$ are 225kJ/mol and 491kJ/mol per site, respectively. Hence, when the sample heats up, the LCC prefers to accept the trial state $[=C=]$ rather than $[-C=]$, $[=C-]$ and $[-C-]$.

We study the stability of the internal LCC as a function of $R_{CNT}$ in Figure 3a. While the chain length is fixed, the optimization process focuses only on the radial direction of the CNT. An isolated long LCC is always unstable in a room-temperature environment due to strong atomic fluctuations [5-7,30-31]. The kink angle is a measure of the atomic fluctuations in the LCC. The kink angle of the LCC in Figure 3b can be used to judge whether the CNT provides an appropriate VDW protection or not. The average kink angle of the internal LCC in (5,1)CNT is only 0.3 degree (or pivot angle = 179.7 degrees) in Figure 3b. This explains why the LCC surrounded by (5,1)CNT has the highest chain stability factor where self-dissociation or chain breakage is less likely. Our simulation result is consistent with the experimental results that the internal LCC in a (6,4)CNT is longer than that in a (6,5)CNT [9].

The chain-stability factor of the LCC@(6,4)CNT increase upon cooling (see Figure 4) because the movement of carbon atoms is inactive at low temperatures [27]. If all covalent bonds in the LCC are stable, the LCC with a length of 15000 atoms should have weaker atomic fluctuations than the LCC with a length of 6000 atoms. However, extensive studies of a LCC show that the elongation of a LCC always leads to chain termination or self-dissociation [4-7,9,30-32,36]. So far, the formation of all stable covalent bonds in an extremely long LCC has not been observed experimentally (48 carbon atoms at most). To make a longer LCC unstable in our model, the Hooke's factor is used to tune the trial range of atomic fluctuations. When our previous LCC model is run this time, the maximum bond length of an isolated LCC calculated by the thermal expansion coefficient is set to $\ell_{c-c} + \ell_{c-c}\alpha\Delta T \sim 1.56$Å manually. In this case, all covalent bonds are stable and eventually $\ell_{carbyne}\big|_{0K}$ is shorter in a longer LCC.

The accuracy of the thermal expansion coefficient in our model was justified by Costa et *al*'s work [29]. After substitution of $\ell_{carbyne}|_{0K}$ into the pre-calibration process, the trial Hooke's factor increases for a longer LCC. By setting the maximum bond length in this carbyne@nanotube model to ~1.73Å [28], we impose a boundary condition that chain breaking is not allowed even if two carbon atoms are ~1.73Å apart. Then our Monte Carlo system can force the atoms to form unstable covalent bonds at any chain length L in excited states. The unstable covalent bonds associated with large atomic fluctuations outweigh the effect of stabilization towards the bulk state. Therefore, the LCC with a length of 15000 atoms (Figure 4b) has a lower chain stability factor than the LCC with a length of 6000 atoms (Figure 4a). Without this boundary condition, we cannot adequately compare the stability of the different LCC@(Nc,Mc)CNT types at the same length.

Although the Hooke's factor is designed to be semi-empirical, our simulation can mimic the stability of a long LCC that is consistent with experimental observations, with the cut-off length (5750-atom-long) in Figure 5 being comparable to the experimental data (~6000-atom-long) [9]. Indeed, the Hooke's factor is just a ratio of velocity, because any elastic force multiplied by $<\delta t>/M$ gives a unit of velocity [27]. Therefore, $v \cdot f(Hooke)$ can be read as an adjustment of the trial velocity. If a carbon chain is going to break [28], the Hooke's factor in a local region approaches 1, so using the free-particle velocity as the maximum velocity at the trial state would be appropriate [27]. The random number $R_c$ refines the spatial fluctuations to aid the process of energy minimization.

In our simulation, a single-walled CNT is used instead of a double-walled CNT to protect the LCC. Using double-walled CNTs should provide better protection for the internal LCC [4,9]. Although our DFT simulation runs a parallel array of LCC@(N$_c$,M$_c$)CNT with a lateral spacing of 0.5nm, the lateral wall-to-wall interaction stabilizes the single-walled CNT in a good shape. The cut-off length of the LCC@(6,4)CNT is slightly underestimated in our model. There are some sources of error, the $\ell_{carbyne@CNT}|_{0K}$ in the Hooke's factor is obtained from the pre-calibration process for N > 4000, but the $K_{carbyne@CNT}$ is the spring constant of an infinitely long LCC. Since the computational cost of DFT simulation may not be realistic when the size exceeds 4000 atoms per unit cell [21], our model assumes that the mechanical properties of LCC consisting of more than 4000 atoms are likely to move towards the bulk state [33,34]. Moreover, the iterative process of our Monte Carlo simulation is not applicable to the atoms in CNTs, as experimental results shows that the atomic vibrations of CNTs are much weaker than those of the internal LCCs. Based on these experimental findings, we can assume that the atoms in the CNTs are relatively at rest. Although the cut-off length is slightly underestimated by these approximations, the computational cost becomes affordable. The inset in Figure 5 confirms that 250000 Monte Carlo Steps are sufficient to drive the system to equilibrium, even if the longest sample is chosen.

The shape of the CNT plays an important role in stabilizing the carbyne@nanotube. The compressive strain changes the CNT from a circular to an elliptical shape. Although the elliptical CNT has a slightly more negative VDW interaction along the minor axis (shorter axis), the VDW interaction along the major axis (longer axis) is much weaker. This is the reason why radial compression destabilizes the LCC@(6,4)CNT. On the other hand, a stronger spatial fluctuation in the internal LCC is observed in porous CNTs. If the selected atom in the internal LCC sums up the negative VDW terms along the angular plane ϕ, the missing carbon atoms in the porous CNT cannot turn the VDW force more negatively, and thus maintaining a good sample quality of CNT is important to prolong the internal LCC.

Our Monte Carlo results are consistent to several experimental observations, such as (1) The LCC in a (6,4)CNT is more stable than the LCC in a (6,5)CNT [9]; (2) The quality of the CNT affects the stability of the internal LCC [4,9]; (3) The cut-off length of the LCC@(6,4)CNT is about ~6000 atoms long [9]; (4) Polyyne is the dominant phase in LCC@(N$_c$,M$_c$)CNT at room-temperature [9]. These Monte Carlo results encourage us to predict the cut-off length of the internal LCC surrounded by different CNT types in Table 4. Based on our Monte Carlo model, the LCC surrounded by a (5,1)CNT should be the longest among the others because the atomic fluctuation is the weakest. The effect of free radical electrons [27] in LCC is ignored in our model, since the probability of forming a polyyne phase is close to 1. When the trial kink angle is large, the selected carbon atom experiences a non-uniform VDW interaction along the angular $\phi$ plane. The non-uniform VDW interaction makes the Hamiltonian more positive, which is consistent with the data in the compressed CNT. For example, the selected carbon atom in the LCC orthogonally shifted by 0.01nm within the (5,4)CNT increases the VDW energy by 2% more positive.

Our model assists in the exploration of the chain-breaking process and the growth mechanism of carbyne. An overlengthen carbyne (exceeding the cut-off length) always breaks into fragments during sample fabrication due to chemical instability [36]. What's happen to an overlengthen carbyne before it breaks naturally? To answer this question, an overlengthen carbyne needs to be generated in our simulation model, and "visualize" the instantaneous atomic fluctuations right before it breaks. A carbon chain made up of 5500-6000 atoms (~780nm) ought to be stable inside (6,4)CNT. When the LCC@(6,4)CNT is compressed by ~4% (Figure 6), the ~780nm-long internal carbyne becomes overlengthened which may be fragmentized. Nair et al [34] justified that a mechanical strain of at least 5% is needed to break a finite-length carbyne. In other words, the carbon chain should break when the $C-C$ bond length reaches 1.5Å x (1+5%) ~1.6Å. After the 6000-atom-long LCC@(6,4)CNT is compressed, the seven sharp kink structures are always connected by the $C-C$ bond lengths above 1.6Å which may initiate chain breakage. Hence, an effective covalent bond between the sites '2' and '3' (Figure 6) should vanish. Reestablishment of an effective covalent bond between the sites '1' and '3' is unlikely to occur because our ab-initio calculation shows that the carbon chain with a pivot angle around 20 degrees generates a local electrostatic repulsion at least two times stronger than the triple-bond energy. In addition, for the case of pivot angle ~ 50 deg (Figure 6), the carbon atom on the site '3' should be unable to form an effective covalent bond with the CNT wall because the radial separation between the carbon atom on the 'site 3' and the CNT wall is still as large as 1.96Å. As a result, the LCC in the compressed (6,4)CNT is probably fragmentized into 8 pieces during the test 1 where the average size of the carbyne fragment is 780nm / (7+1) = 97nm. The eight short carbyne chains should consist of 103-1=102 atoms, 576-103=473 atoms, 932-576=356 atoms, 1853-932=921 atoms, 4318-1853=2465 atoms, 4943-4318=625 atoms, 7505-4943=762 atoms, 6000-5705=295 atoms, respectively. Our model can be used to forecast where the long carbon chain should break at finite temperatures [36], and the instantaneous atomic configuration of carbyne right before it breaks. Our model does not probe the further atomic movement of carbyne fragments after the long chain breaks. Based on ref 33, the mechanical properties of carbyne shows size effect up to N ~ 20. If the carbyne fragment is less than 20-atom-long, the atomic spring constant in the Hooke's factor should be revised. However, the shortest carbyne fragment formed by compression still contains over 100 atoms and hence the Hooke's factor is still applicable to monitor the fragmentation of carbyne.

As an example, we investigate if scientists want to use LCC@(6,5)CNT in the transistor for emerging continuously tunable band gap. Experiments show that long carbyne enclosed by metallic (6,5)CNT do exist [9] and the precise control of chiral angles in CNT enters a new era [35]. When LCC@(6,5)CNT is unstrained, the cut-off length of LCC is ~720 nm. After LCC@(6,5)CNT is strained by 1% longitudinally, the cut-off length drops to ~600 nm. To ensure the internal carbyne stable during the operation, the maximum length of the unstrained LCC@(6,5)CNT should not exceed ~600 nm. The ~600 nm long carbyne in the (6,5)CNT is stable at 1% strain. When the LCC@(6,5)CNT is unstrained, the stability of the LCC is even stronger based on the analysis of Figure 5. Hence, neither 1% strain nor unstrained condition destroys the internal carbyne, and more encouragingly, the 1% strain should change the band gap of the carbyne by ~15% continuously [6] which provide a precise control in the energy barriers in the transistors

The low computational cost of our Monte Carlo model is credited to two approximations. The first approximation is that the atomic positions of the nanoreactor (finite-length CNT) in our Monte Carlo model is obtained by scaling the repeating unit in the ab-initio calculations. This speeds up the computational progress because a time-consuming geometric optimization process of a big supercell (over 100000 atoms in a non-repeating unit) is avoidable. To validate this approximation, we consider the length dependence of the Raman spectrum of a finite-length CNT where the size effect starts to pale for a CNT length longer than ~200nm [37]. As the shape of the Raman spectrum is closely related to the atomic positions of materials and the minimum length of the nanoreactor CNT in our model is far above 200nm, it is arguably that the first approximation will not create a large error in Monte Carlo results. The second approximation is that the Monte Carlo iterations are conducted to the internal chain only. This approximation can be validated by the experimental fact that the atomic fluctuations of CNT are relatively at rest when compared to the internal chain [9]. With the second approximation, the size of one-dimensional array in our Monte Carlo simulation is 4000-15000 only, which reduces the computational time massively without creating a large error in Monte Carlo results.

## 5. Conclusions:

Our Monte Carlo model can be used to predict the cut-off length of the LCC encapsulated by a CNT at any temperature. Synthesis of the internal carbon chain with more than 6000 atoms is possible if the temperature and the type of CNT are chosen correctly. The radius, chirality and quality of CNT have a great influence on the stabilization of the internal LCC. Radial compression on the carbyne@nanotube destabilizes the internal LCC. Apart from these, our investigation of the compressed carbyne@nanotube leads to discover the mechanisms behind the chain-breaking process and the probable formation of carbyne fragments. Our model opens a new way to study the stability of carbyne in the size of 4000-15000 atoms at finite temperature, which is a steppingstone for experimental bottom-up studies of the edge magnetism from the monoatomic structure to a higher-dimensional Bravais lattice and continuously tunable energy barriers in the CNT-based transistors.

**Acknowledgements:**

We appreciate the Research Institute for Advanced Manufacturing and the Industrial Centre at The Hong Kong Polytechnic University to support this project. We thank the Shanghai Bangtu Culture Media Co.,Ltd to assist in graphical productions. C.H.Wong thanks PolyU (UGC) for the research funding (1-45-37-BD5C). L.Shi. thanks the National Natural Science Foundation of China (51902353) and Fundamental Research Funds for the Central Universities, Sun Yat-sen University (22lgqb03).

**Author contributions:**

CH.Wong planned this project. CH.Wong designed Monte Carlo algorithms, implemented the ab-initio simulations and analysed results. CH.Wong, R.Lortz, L.Shi, CW Leung, CL Mak discussed data. CH.Wong, R.Lortz, YM.Yeung, L Shi wrote manuscript. CHWong, X.Zhao, CY.Tang and WC.Law handled graphics.

**Competing Interests**

The authors declare no conflicts of interest.

**Data Availability Statement**

Data availability is possible upon a reasonable request